# Design and Optimization of Low Energy Beam Transport for TAC Proton Facility

A. Caliskan[1], *H. F. Kisoglu[2], S. Sultansoy[3,4], M. Yilmaz[5]

[1]Department of Engineering Physics, Gumushane University, Gumushane, Turkey
[2]Department of Physics, Aksaray University, Aksaray, Turkey
[3]Institute of Physics, Academy of Sciences, Baku, Azerbaijan
[4]Physics Division, TOBB University of Economics and Technology, Ankara, Turkey
[5]Department of Physics, Gazi University, Ankara, Turkey

**Abstract**

In this study, a low energy beam transport (LEBT) channel for the proton linac section of the Turkish Accelerator Center (TAC) has been designed by using TRAVEL code. Commonly used LEBT including two focusing solenoid magnets will transport and match the $H^-$ beam from a volume source to RFQ. In the beam dynamics simulations of such a LEBT line, 95% space-charge compensation (SCC) has been considered in this study. We aimed to find out the determination of our RFQ input parameters that gives the best possible beam quality at the entrance of the RFQ using beam collimator in the LEBT line as an alternative way. In this way, we have acquired the best possible beam quality on RFQ input plane as well as optimizing the LEBT line.

## 1. Introduction

The Turkish Accelerator Center (TAC) is a regional project [1] and has been developed with support of the Turkish State Planning Organization (DPT) by collaboration of several Turkish universities. Its conceptual design report was completed in 2005 and technical design report (TDR) studies have been continued since 2006. Today, TAC project includes linac-ring type super charm factory, synchrotron light source based on positron ring, free electron laser based on electron linac, GeV scale proton accelerator and TAC test facility [2].

The proton accelerator construction will have 3 MeV, 100 MeV and 1 GeV phases. It will give an opportunity to produce secondary muon and neutron beams for applied research fields in addition to primary proton beam. In the muon region, a lot of applied investigations in some research areas like high-Tc superconductivity, phase transitions, impurities in semiconductors will be performed using powerful muon spin resonance (µSR) method [3]. In the neutron region, it is planned to be used for different fields of science such as engineering, molecular biology and fundamental physics. Additionally, accelerator driven systems (ADS) technology application of GeV energy proton accelerator becomes very attractive for our country since Turkey has essential thorium reserves [4].

The proposed proton accelerator is linear and its fundamental accelerator structures are an RF volume ion source, low energy beam transport (LEBT) channel, RFQ (radio-frequency quadrupole) and medium energy beam transport (MEBT) channel for the low (3 MeV) energy section. For medium energy (100 MeV) section, we have planned using drift tube linac (DTL) [5, 6] and coupled cavity drift tube linac (CCDTL) cavities. For high energy (1 GeV) section, using normal conducting or super conducting cavities is under investigation.

*Corresponding author: hasanfatihk@aksaray.edu.tr

In this paper, we present the results of the beam dynamics simulations for designing of LEBT part of the TAC proton linac in alternative way. Main components and parameters of the LEBT channel are mentioned in the next section. In the third section, beam dynamics studies in which we have determined the magnetic field values of solenoid magnets have been performed. To increase the quality of the beam we have performed a collimator study in the section 4 in accordance with main objective of our methodology. Finally, we summarized our results in section 5.

## 2. Low Energy Beam Transport Line

The low energy beam transport (LEBT) system of the TAC proton linac section will be located between an ion source and the RFQ. The LEBT channel transports and matches the beam from the source to the RFQ plane. It consists of some diagnostics elements such as faraday cup, BC transformers, etc. as well. Similar proton linac projects mostly use two solenoids for the LEBT while there are either electrostatic (ES) or magnetostatic focusing configurations. It is known that solenoid magnet is more advantageous than a quadrupole doublet as it focuses on both of transverse planes simultaneously. Configuration with two solenoid magnets is planned to be used for the LEBT of the TAC proton linac.

TRAVEL [7] beam dynamics simulation code, which includes space-charge effects for both bunched and dc beams, has been used to design such a LEBT line and study beam dynamics. In the beam dynamics simulations, we have taken into account the effect of the space charge compensation method. During these simulations $H^-$ beam extracted from an ion source was transported to the best possible RFQ input parameters for which the high beam quality can be obtained. The parameters of such a LEBT line, intended for TAC proton accelerator, are given in Table 1.

**Table 1** Proposal LEBT line parameters for TAC proton facility

| Parameters | Length (mm) |
|---|---|
| Drift | 200 |
| Solenoid | 300 |
| Drift | 900 |
| Solenoid | 300 |
| Drift | 250 |
| Beam aperture | 55 |

As is seen from Figure 1, the LEBT line consists of two solenoids and three drifts. Aperture size of vacuum tube in which the beam goes through is 5.5 cm along the whole LEBT and it is equal to the radius of the solenoid.

*Corresponding author: hasanfatihk@aksaray.edu.tr

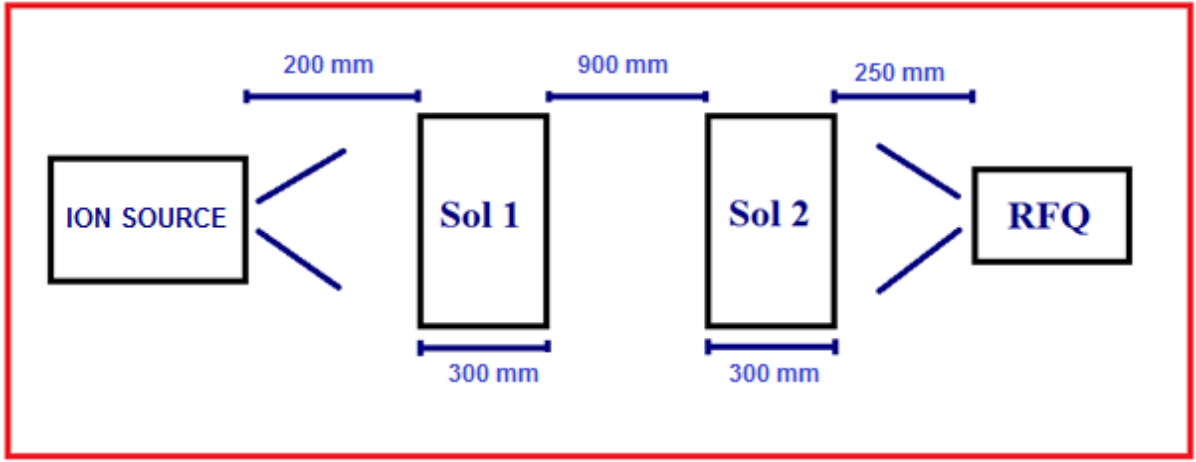

**Figure 1.** The block diagram of proposed LEBT channel for TAC proton facility

## 3. Beam Dynamics Simulations

We have suggested that H⁻ ion beam is extracted with 80 mA current and 80 keV kinetic energy from a volume source. The initial parameters of such a beam were assumed to be $\varepsilon_{xx'} = \varepsilon_{yy'} = 0.37 \times 10^{-6}\,\text{m·rad}$ (normalized, RMS), $\beta_{xx'} = \beta_{yy'} = 0.33$ m/rad and $\alpha_{xx'} = \alpha_{yy'} = -6.2$. The beam was passed through the LEBT line and evolutions of parameters were analyzed by using TRAVEL code. Furthermore, the effects of space-charge forces were regarded and 95% space-charge compensation (SCC) was considered to minimize these effects during the simulations. We aimed at increasing the beam brightness as much as possible at the end of the design as it is a measure of the beam quality. Furthermore, halo parameter at the entrance of the RFQ has been kept in view as another condition. So, we tried to enhance the particle density for the subsequent parts of the linac and to weaken the machine activation on the cavity walls.

We used the Linac4 [8] solenoids in beam dynamics simulations. The strengths of both magnets are 3.20 kGauss. We tuned the magnetic fields, variating the scaling factors, to acquire minimum emittance change and maximum beam transmission. The field strengths of roughly 3.0 kGauss and 2.6 kGauss, respectively, were obtained from tuning (Figure 2).

*Corresponding author: hasanfatihk@aksaray.edu.tr

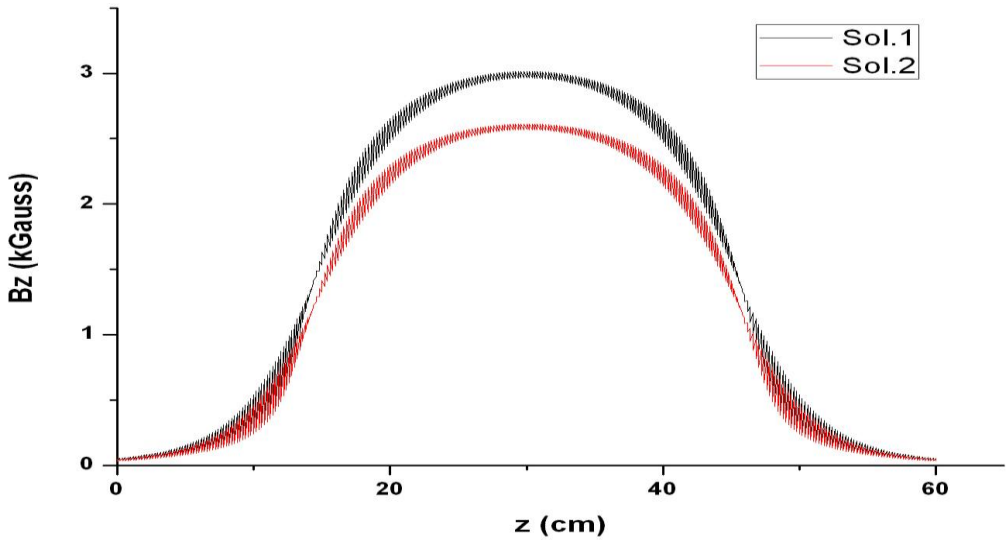

**Figure 2** Pattern along the beam axis of the axial magnetic field that gives minimum emittance change and maximum transmission.

The H⁻ beam with the initial parameters was passed through the LEBT line without collimator and evolutions of the beam size and emittance of the beam have been obtained as shown in Figure 3.

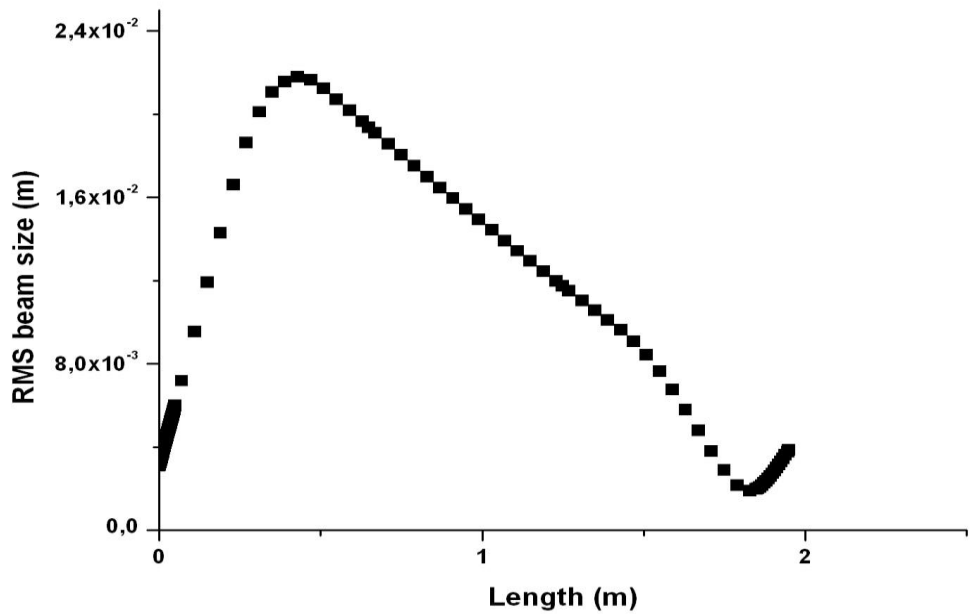

(a)

*Corresponding author: hasanfatihk@aksaray.edu.tr

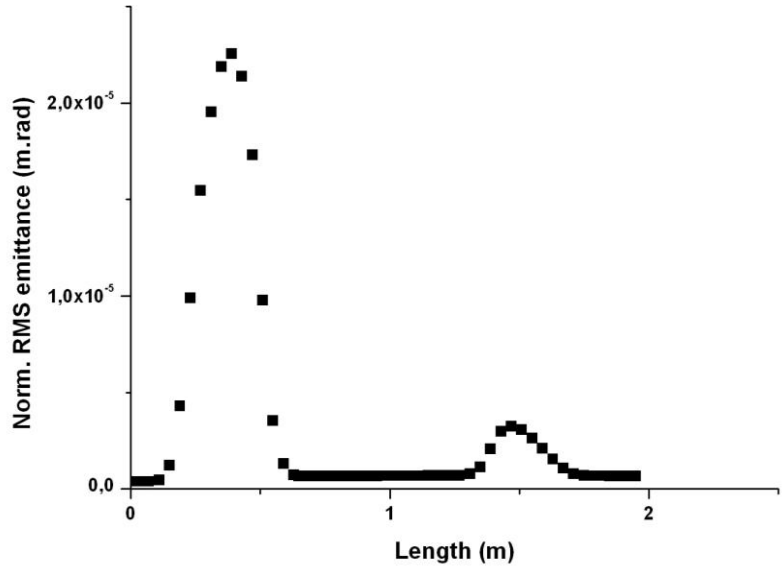

(b)

**Figure 3** Beam size(a) and emittance(b) evolutions without collimator.

As is seen from Fig.3, there are quite differences between initial and final values. The change especially on the emittance is roughly 2 fold. The growth on the emittance and beam size manifest itself especially at the first drift, up to the first solenoid. This growth effect the evolution of the emittance, halo formation and brightness along the LEBT. Because the higher the divergence when entering the solenoid, the higher the emittance growth when exiting the solenoid[9]. Nevertheless, we obtained a transmission of 96.4% without collimator usage as shown in Figure 4. It can be also seen from Fig. 4 that the particle lost mostly occurs in the first solenoid.

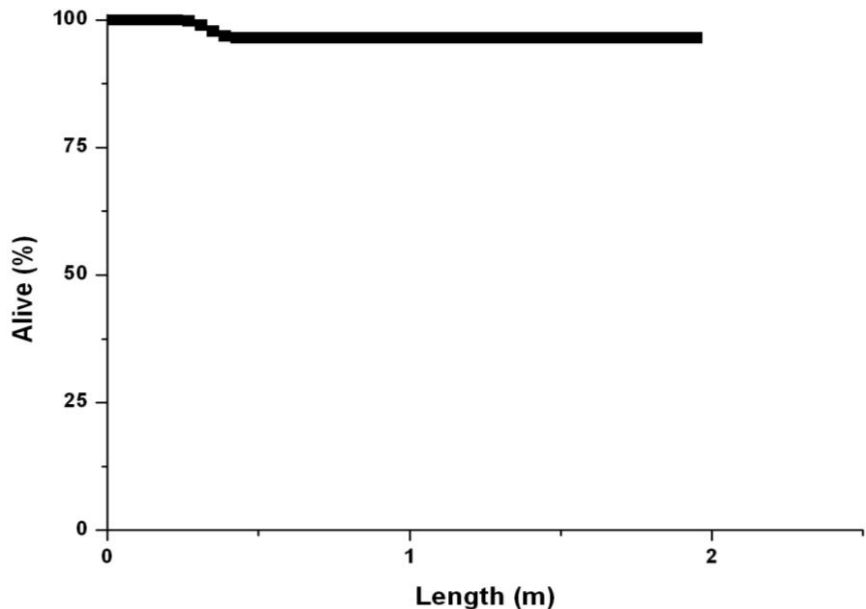

**Figure 4** Beam transmission along the whole LEBT line

*Corresponding author: hasanfatihk@aksaray.edu.tr

## 4. The Collimator Study

A beam collimator, which has smaller aperture size than that of vacuum tube, can be used at drift regions in the LEBT line[10]. Thus, if we use the collimator at the first drift, we can avoid unwanted particles that cause to increase emittance and beam size. Furthermore, brightness of the beam can be increased under favour of this method. So, the beam can be transported to next parts intensively. On this basis, we placed a collimator of 1 cm along the beam axis at the first drift where there is no particle lost. We have used the collimator only at the first drift, as the reason of the overall growth on the emittance and halo formation along the LEBT line could be this increment at the first drift.

First of concerning with collimator, we determined the position of the collimator at the first drift. Beam transmission and halo formation were taken into account during this tuning. We located the collimator 1 cm, 2 cm, 3 cm and 3.5 cm, respectively, away from the beginning of the first drift of the LEBT on the beam axis. For four cases, the aperture size of the collimator is a provisional value of 0.3 cm. The changing on particle losses versus variation of position of the collimator can be seen on Figure 5.

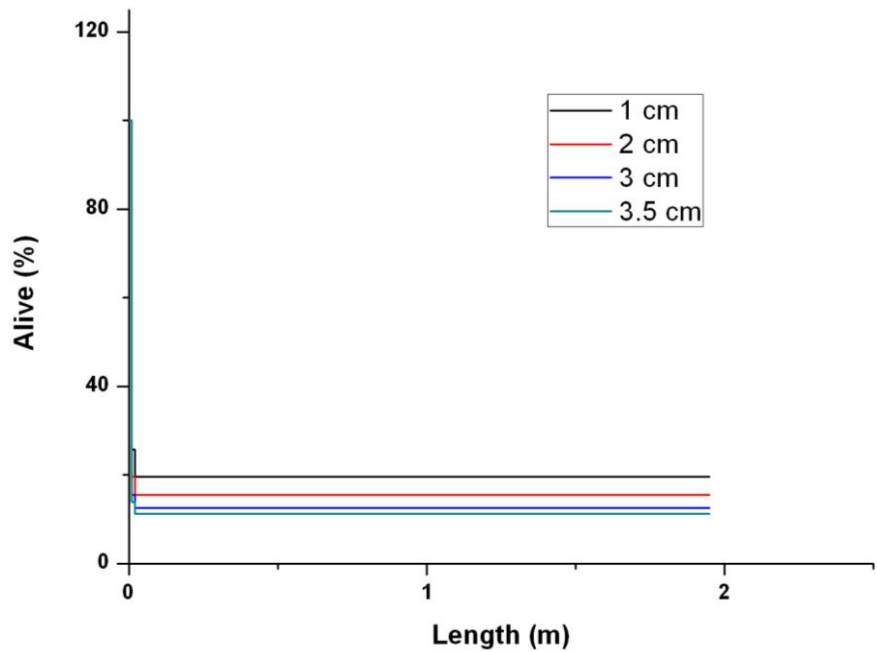

**Figure 5** Beam transmission along the whole LEBT under the influence of collimator position at first drift.

*Corresponding author: hasanfatihk@aksaray.edu.tr

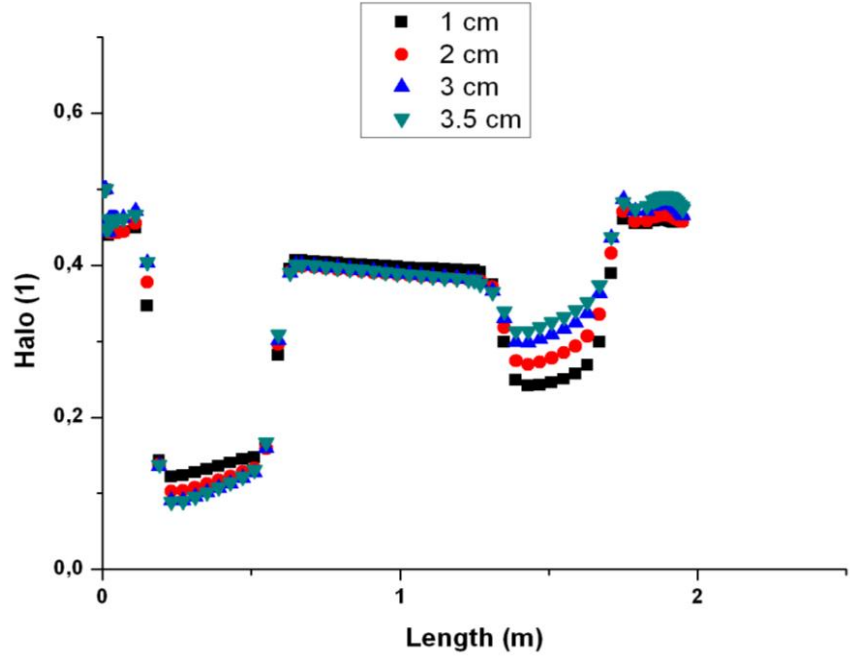

**Figure 6** Beam halo formation along the whole LEBT under the influence of collimator position at first drift.

Another parameter we regarded is, beam halo, the formation of a low density halo surrounding the beam core. This is due to space-charge induced emittance growth and can cause beam loss [11]. However, it gives a measure of consequent machine activation [9]. Halo intensity parameter in the $i\,th$ plane, $H_i$, for a continuous beam is [12];

$$H_i = \frac{\sqrt{3I_4^i}}{2I_2^i} - 2 \qquad (1)$$

where $I_2^i \equiv <q_i^2><p_i^2> - <q_i p_i>^2$, $I_4^i \equiv <q_i^4><p_i^4> +3<q_i^2 p_i^2>^2 -4<q_i p_i^3><q_i^3 p_i>$ are the second and the fourth momenta and $q_i$, $p_i$ are the position and momentum, respectively, in the $i\,th$ plane. Halo formation along the LEBT line versus variation on the collimator position has been given on Figure 6. We placed the collimator at 1 cm from the beginning of first drift regarding to Fig.5 and Fig.6 which are trivial results that describe beam size increases when distance between the ion source and collimator increases. But 1 cm, according to this result, is the sufficient enough to show that collimator usage effects the emittance growth and halo formation, thus, beam brightness.

After the collimator position, we attempted to determine the precise beam aperture of collimator. Brightness of the beam was the figure of merit while tuning the aperture. The variation on beam brightness under the influence of aperture size is as shown in Figure 7. According to Fig.7 we can say that beam current passing through the collimator increases rapidly compared to emittance as the beam aperture enlarges up to a maximum point and after this point all of the particles in the beam flow through the collimator and beam current is fixed while the emittance continues to grow. After a while emittance and brightness are pegged.

*Corresponding author: hasanfatihk@aksaray.edu.tr

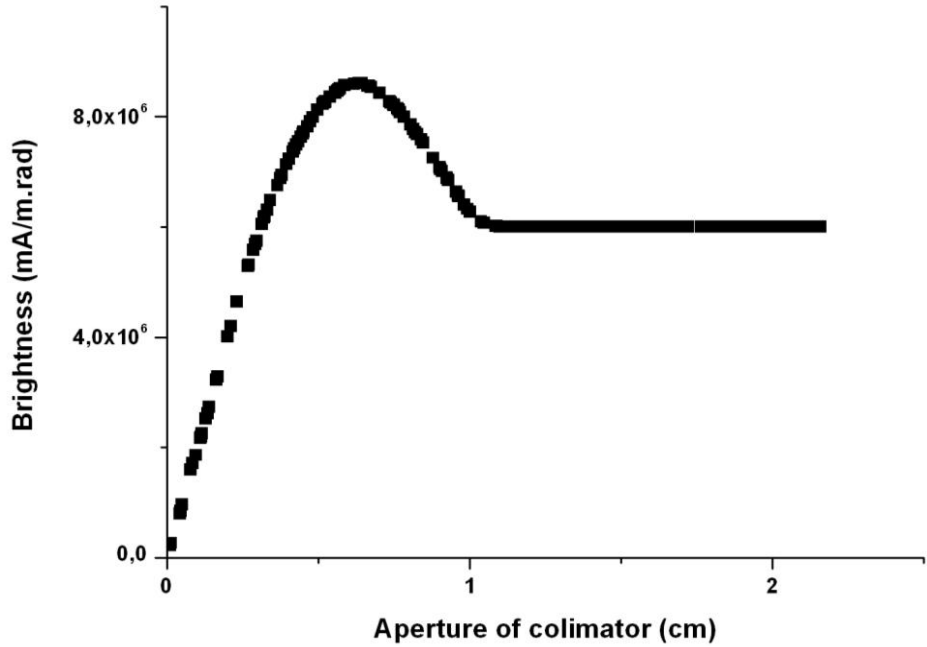

**Figure 7** Beam brightness variation versus aperture size of collimator.

We chose 0.636 cm as aperture size of the collimator according to Fig.7. The beam transmission is roughly 65% in case of collimator usage. However, the beam emittance and halo formation were decreased by 52.9% and 56.5%, respectively. Furthermore, the beam brightness was increased by 43.6%. A transmission value of 65% is satisfactory enough because of the necessity of different applications. This transmission corresponds to beam current of 2.6 mA for non-compensated 5% portion, i.e. 4 mA, of the whole beam. If we keep in mind the compensated 95% portion, i.e 76 mA, we obtain a beam current of 78.6 mA at the exit of the LEBT. In conclusion, we have a transmission of ~98.2% at the exit of the LEBT line. In a similar vine, the transmission for the LEBT line without collimator usage is ~99.8%.

The comparison of beam size and emittance, belong to LEBT with/without collimator, is as shown in Figure 8. Moreover, the RFQ input parameters for both LEBT line have been summarized in Table 2 according to simulation results.

*Corresponding author: hasanfatihk@aksaray.edu.tr

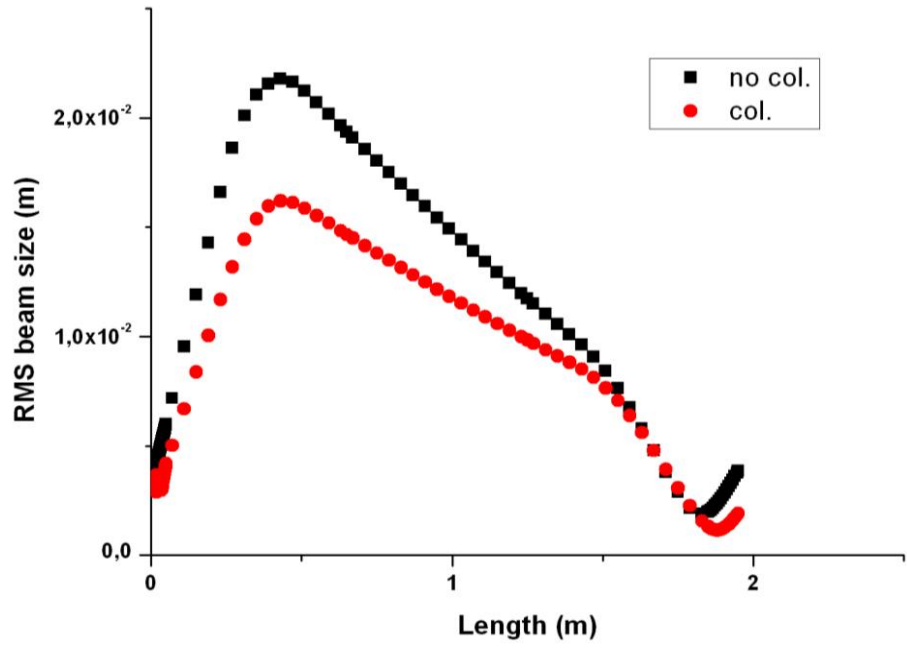

(a)

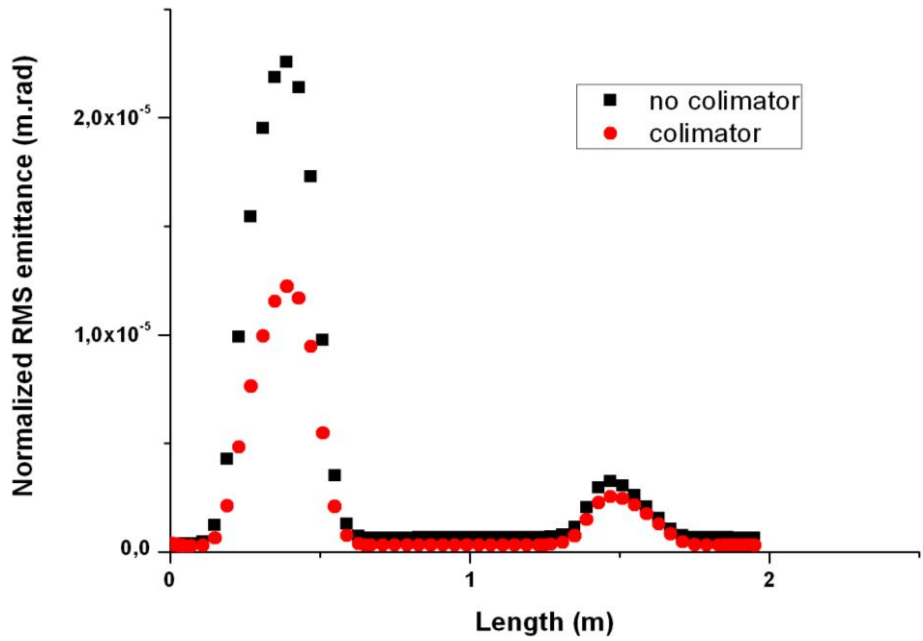

(b)

**Figure 8** Beam size(a) and emittance(b) comparison in case of collimator usage and no collimator usage.

**Table 2** The RFQ input parameters with/without collimator.

|  | Transmission (%) | Norm. RMS Emittance (m.rad) | Halo (1) | Brightness (mA/m.rad) | $\beta$ (m/rad) | $\alpha$ |
|---|---|---|---|---|---|---|
| **No collimator** | 99.8 | $0.64 \times 10^{-6}$ | 1.33 | $5.99 \times 10^6$ | 0.30 | -1.89 |
| **Collimator** | 98.2 | $0.30 \times 10^{-6}$ | 0.58 | $8.60 \times 10^6$ | 0.16 | -1.44 |

*Corresponding author: hasanfatihk@aksaray.edu.tr

## 5. Conclusions

Before this study, accelerator parts (DTL, CCDTL, and CCL) of proton linac of the TAC project was designed [5, 6, 13]. The design studies of the injector part of proton linac are ongoing. It is planned to have a ring after the linac, in TAC, for the time being. So, an RF volume ion source will be used in this case. A microwave-off resonance or an RF ion source which are used to obtain $H^+$ ion beam could be used, if it is considered to linear. The second component of injector part is low energy beam transport (LEBT) system. We insist on the LEBT configuration that includes two solenoids. The extensive studies on the layout of LEBT including diagnostics such as faraday cup, SEMs, BC transformers, etc. are under investigation. Radio frequency quadrupole (RFQ) is the other sub-system of the injector part of proton accelerator. In this part, the beam will be accelerated up to 3 MeV.

We have done this study for finding out a novel method for determination of the RFQ parameters. For this purpose, a LEBT channel that consists two solenoid magnets has been designed and optimized in an alternative way. We have acquired the best possible beam quality on RFQ input plane as well as optimizing the LEBT line using this method. $H^-$ ion beam, extracted with 80 mA current and 80 keV kinetic energy from a volume source, was used in simulations. In these beam dynamics simulations, TRAVEL code was used regarding to effects of space-charge forces. For minimizing these effects 95% space-charge compensation (SCC) was considered. We aimed at increasing the beam brightness as much as possible at the end of the design as it is a measure of the beam quality. Furthermore, halo parameter at the entrance of the RFQ has been kept in view as another condition. So, we tried to enhance the particle density for the subsequent parts of the linac and to weaken the machine activation on the cavity walls. This goal can be actualized by increasing the aperture size of the solenoids and drifts enhancing the transmission. However, the larger beam aperture we choose, the more expenditure we get as the cost of LEBT is proportional roughly $r_0^2$ where $r_0$ is the beam aperture[14]. Furthermore, focusing on the beam axis will be smoother if we increase the radius of the solenoids. So, there are some restrictions on aperture size. Also, all factors that enhance the halo formation and emittance growth along the LEBT line should be minimized as much as possible. For this purpose, we have performed a collimator study for 95% SCC and seen that beam quality increased although particle loss occurred. However, beam brightness was increased roughly by 2 fold and the halo formation was decreased by half. The increment of brightness is a result of decreasing of the emittance. Furthermore, there is quite small difference in transmission values between with/without collimator usage.

## Acknowledgements

We would like to thank Turkish State Planning Organization (DPT), under the grants no DPT-2006K120470, and Scientific and Technological Research Council of Turkey (TUBITAK) for their supports. We also special thanks to Alessandra Lombardi for helps.

*Corresponding author: hasanfatihk@aksaray.edu.tr

*Corresponding author: hasanfatihk@aksaray.edu.tr